# Statistical Delay Control and QoS-Driven Power Allocation Over Two-Hop Wireless Relay Links

Qinghe Du[†], Yi Huang[‡], Pinyi Ren[†], and Chao Zhang[†]

[†] Department of Information and Communications Engineering, Xi'an Jiaotong University, China
[‡] Qualcomm Research Center, San Diego, USA
E-mail: {duqinghe@mail.xjtu.edu.cn, yihuang@ieee.org, pyren@mail.xjtu.edu.cn, chaozhang@mail.xjtu.edu.cn}

*Abstract*— The time-varying feature of wireless channels usually makes the hard delay bound for data transmissions unrealistic to guarantee. In contrast, the *statistically-bounded delay with a small violation probability* has been widely used for delay quality-of-service (QoS) characterization and evaluation. While existing research mainly focused on the statistical-delay control in single-hop links, in this paper we propose the QoS-driven power-allocation scheme over two-hop wireless relay links to statistically upper-bound the *end-to-end* delay under the decode-and-forward (DF) relay transmissions. Specifically, by applying the effective capacity and effective bandwidth theories, we first analyze the delay-bound violation probability over two tops each with independent service processes. Then, we show that an efficient approach for statistical-delay guarantees is to make the delay distributions of both hops *identical*, which, however, needs to be obtained through *asymmetric* resource allocations over the two hops. Motivated by this fact, we formulate and solve an optimization problem aiming at minimizing the average power consumptions to satisfy the specified end-to-end delay-bound violation probability over two-hop relay links. Also conducted is a set of simulations results to show the impact of the QoS requirements, traffic load, and position of the relay node on the power allocation under our proposed optimal scheme.

*Index Terms*— Quality-of-service (QoS), delay, effective capacity, effective bandwidth, delay-bound violation probability, fading channels, relay, multi-hop.

## I. INTRODUCTION

PROVISIONING of multimedia services has been widely recognized as one of the major features in next-generation wireless networks. Since most multimedia applications are delay sensitive, conventional research aiming at the maximization of network throughput already cannot catch up with the rapidly increasing demands on multimedia services. Consequently, delay quality-of-service (QoS) provisioning over wireless channels has attracted more and more research attention [1]–[7], [9]–[12], [14]–[16], [20].

There are mainly three approaches for delay-QoS guarantees over wireless channels proposed by existing research works. The most frequently used approach is to guarantee a minimum required transmission rate through resource allocation. Clearly, maintaining a specified transmission rate requires the power control to compensate the deep channel fading. It is well-known that in Rayleigh fading channels with upper-bounded power budget, the transmission rate which can be guaranteed is equal to zero. As a result, such a requirement often cannot be satisfied. The other way for delay-QoS provisioning is to strictly control the delay under a specified threshold. However, the time-varying feature of wireless channels usually makes the hard delay bound for data transmissions unrealistic to guarantee. In contrast, statistical delay-QoS provisioning has gradually replaced the deterministic delay-QoS provisioning (e.g., hard delay bound) for delay-sensitive services.

One of the typical metrics for statistical delay-QoS provisioning is the delay-bound violation probability, which has been investigated in several research works. In [14], the authors built the effective capacity framework to evaluate and characterize the delay-QoS performances for a single wireless link. In [7], the authors proposed the delay-QoS driven base-station selection algorithm to satisfy multiple downlink users' specified delay-bound violation probabilities. In [16], the authors developed the time-slot allocation schemes for multi-layer video unicast/multicast over wireless channels transmissions to control the delay-bound violation probability. However, these works mainly concentrated the single-hope wireless links, which cannot be applied to multi-hop wireless transmissions.

Literatures [1], [2] investigated the delay-constrained relay links in various scenarios, while these works in fact formulate the relay transmission as a single-hop problem. In [5], the authors proposed a modeling technique to characterize the delay-bound violation probability in wired multi-hop mesh networks, and developed the corresponding resource allocation schemes to minimize the video distortion while keeping fairness across multiple connections. This result is hard to be applied to wireless networks, because it is not aware of the fundamental fading feature of wireless channels. In [9], the author gave a relatively loose upper-bound of the delay-bound violation probability for multi-hop data transmissions. In [6], the authors derived a tight lower-bound of delay-bound violation probability for multi-hop mesh networks. However, the above research works do not answer how to efficiently satisfy the specified delay-bound violation probability requirement through QoS-driven resource allocation, which still remains a widely cited open problem.

The research reported in this paper (correspondence author: Pinyi Ren) was supported in part by the National Science Foundation of China under Grant No. 60832007 and the National Hi-Tech Research and Development Plan of China under Grant No. 2009AA011801.

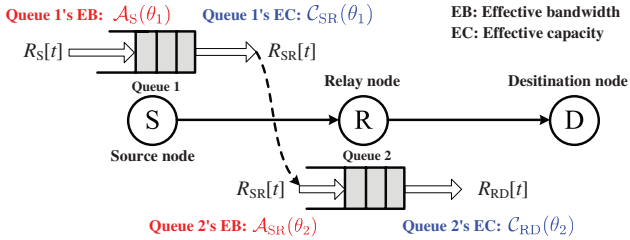

Fig. 1. Queuing model for the two-hop wireless relay links.

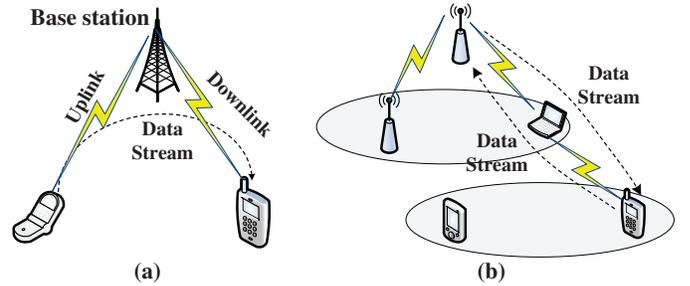

Fig. 2. Typical network structures with two-hop relay links. (a) Cellular networks. (b) Two-tier structure in wireless sensor networks or ad hoc networks.

To overcome the above problem, we propose the QoS-driven power-allocation scheme over two-hop wireless relay links in order to statistically upper-bound the *end-to-end* delay, where the relay transmissions employ the decode-and-forward (DF) mode. The two-hop relay links are widely used in various types of network structures, such as cellular networks and two-tier wireless networks. By applying the effective capacity and effective bandwidth theories, we first analyze the delay-bound violation probability over two tops each with independent service processes. Then, we show that an efficient approach for statistical-delay guarantees is to make the delay distributions of both hops *identical*, which, however, needs to be obtained through *asymmetrically* resource allocations over the two hops. We further formulate and solve an optimization problem to minimize the average power consumptions to satisfy the specified end-to-end delay-bound violation probability in two-hop relay links. We also present simulation results to demonstrate the effectiveness of our proposed QoS-driven power allocation scheme.

The rest of the paper is organized as follows. Section II describes the system model for the two-hop wireless relay links. Section III analyses the delay-bound violation probabilities over the two-hop wireless relay links. Based on the analyses in Section III, Section IV formulate the optimization framework to minimize the total power used for two-hop relay links subject to the statistical delay-QoS requirements. Section V presents the numerical results for performance analyses. The paper concludes with Section VI.

## II. SYSTEM MODEL

### A. System Descriptions

We concentrate on the two-hop wireless relay link, which includes a source node, relay node, and a destination node, as shown in Fig. 1. The source-to-relay and relay-to-destination distances are equal to $d_1$ and $d_2$, respectively. The source node cannot reach the destination node directly. Therefore, the source node first transmits the data to the relay node. The relay node decodes the received signal and then forward the re-encoded data to the destination node. Note that the two-hop relay links are widely applied in various types of wireless networks. For example, in cellular networks depicted in Fig. 2(a), the communications between two end users in a cell requires the source user transmit to the base station first. Then, the base station will decode and forward the data to the destination user. The other typical example is the two-tier network structures as shown in Fig. 2(a), which is often employed in wireless sensor networks and ad hoc networks. In such cases, the communications between a node in the top level and that in the bottom level in fact build a two-hop wireless relay link.

The power gain of the channel between the source-to-relay (SR) link and that between the relay-to-destination (RD) link are denoted by $h_1$ and $h_2$, respectively, where $h_1$ and $h_2$ are independent random variables, which follows the Rayleigh fading model. The channel gains $\{h_i\}_{i=1}^2$ also follow independent fading model, where $\{h_i\}_{i=1}^2$ remain unchanged within a time-frame with fixed length equal to $T_f$, but vary independently from frame to frame. The power of the additive Gaussian noise is set equal to 1.

We consider both the full-duplex mode and the half-duplex mode for relay transmissions. For the full-duplex mode, the relay node can transmit and send data in the same time, corresponding to the case of a powerful relay node with two sets of transceivers, such as the base station in cellular networks. For the half-duplex mode, we suppose that the SR link and RD link each use half of the time frame for transmissions. We then can use a unified model for both the full-duplex and half-duplex modes, where we set the signal bandwidths for both the SR link and RD link equal to B, and set the transmission time for both links equal to T. As a result, for the half-duplex mode we have $T = T_f$ and total bandwidth consumptions are B; for the full-duplex mode we have $T = T_f/2$ and the total bandwidth consumptions are 2B.

### B. Delay QoS Requirements

As discussed in Section I, we in this paper use the statistical metric to characterize the delay QoS requirements. In particular, we require that the end-to-end (source-to-designation) delay-bound violation probability satisfy:

$$\Pr\{D_{\text{total}} > D_{\text{th}}\} \leq \xi, \quad (1)$$

where $D_{\text{total}}$ denotes the end-to-end queueing delay, $D_{\text{th}}$ is the tolerable delay bound, and the $\xi$ is the maximum tolerable delay-bound violation probability. Note that in this paper we mainly focus on the queueing delay caused by the wireless transmissions, since the wireless channels are typical the bottleneck in network transmissions.

As depicted in Fig. 1, both the source and relay node will maintain a queue to buffer the arrival data. The data arrival process for the queue at the source node is denoted by $R_S[t]$ (nats/frame) and the service process is denoted by $R_{SR}[t]$ (nats/frame), where $t = 1, 2, \ldots$, is the time-frame index. Clearly, $R_{RD}[t]$ will also serve as the arrival process of the queue at the relay node. The service process of the queue at the relay node is denoted by $R_{RD}[t]$. Then, the total delay $D_{total}$ is the sum of the queueing delays over these two queues. We in this paper mainly concentrate on the capability of the two-hop wireless relay channels in support delay QoS, and thus assume $R_S[t] = \bar{A}$ is a constant process.

*C. Rate Adaptation*

Both of the source node and the relay node use constant power $\kappa_1$ and $\kappa_2$, respectively, for data transmissions. We assume that the source node knows $h_1$ and the relay node knows $h_2$ in each time frame. Given the channel gains $h_1$ and $h_2$, the service processes $R_{SR}[t]$ and $R_{RD}[t]$ are set equal to the corresponding Shannon capacity as follows:

$$
\begin{aligned}
R_{SR} &= BT \log(1 + \kappa_1 h_1); \\
R_{RD} &= BT \log(1 + \kappa_2 h_2).
\end{aligned} \quad (2)
$$

### III. DELAY-BOUND VIOLATION PROBABILITY ANALYSES

*A. Background on Statistical Delay Analyses for the Single Queue*

Consider a stable dynamic discrete-time queueing system with stationary arrival and service processes. The data arrival-rate and data service-rate of the queueing system are denoted by $C[t]$ (nats/frame) and $R[t]$ (nats/frame), respectively. The $C[t]$ and $R[t]$ change from frame to frame and thus can be characterized as the time-varying processes. Based on the large deviation principal, asymptotic analyses [13] show that

$$
-\lim_{Q \to \infty} \frac{\log(\Pr\{Q > Q_{th}\})}{Q_{th}} = \theta, \quad (3)
$$

for a certain $\theta > 0$, where $Q$ denotes the queue length. Correspondingly, the probability that $Q$ exceeding the given bound $Q_{th}$ can be approximated by

$$
\Pr\{Q > Q_{th}\} \approx e^{-\theta Q_{th}}. \quad (4)
$$

In Eqs. (3)-(4), $\theta > 0$ is a constant called QoS exponent, which is jointly determined by the features of arrival and service processes.

In particular, the effective bandwidth [13] of the arrival process under the given QoS exponent $\theta$ is the *minimum* constant service process required to result in this $\theta$, which is denoted by $A(\theta)$. In contrast, the effective capacity [14] of the service process under the given QoS exponent $\theta$ is the *maximum* constant arrival process that can be supported to obtain this $\theta$, which is denoted by $C(\theta)$. If $C[t]$ and $A[t]$ are time-uncorrelated, [13], [14] show that

$$
\begin{aligned}
A(\theta) &= \frac{1}{\theta} \log E\left[e^{\theta C[t]}\right] \\
C(\theta) &= -\frac{1}{\theta} \log E\left[e^{-\theta R[t]}\right]
\end{aligned} \quad (5)
$$

The authors of [13] further showed that for a dynamic queueing system satisfying Eq. (4) under certain $\theta$, we have

$$
A(\theta) = C(\theta). \quad (6)
$$

Accordingly, the delay-bound violation probability can be approximated by

$$
\Pr\{D > D_{th}\} \approx e^{-\theta A(\theta) D_{th}}, \quad (7)
$$

where $D$ and $D_{th}$ denote the queueing delay and delay bound, respectively.

*B. Statistical Delay Analyses for the Two-Hop Transmissions*

In this section, our target is to derive the complement cumulative distribution function (CCDF) $\Pr\{D_{total} > x\}$ of the end-to-end delay $D_{total} = D_1 + D_2$, where $D_1$ and $D_2$ are the queueing delays of the SR and RD links, respectively.

Suppose that both queues for the SR and RD links are stable. Then, based on Eq. (6), there exists a certain $\theta_1$ such that the effective bandwidth of $R_S[t]$, denoted by $A_S(\theta_1)$, is equal to the effective capacity of $R_{SR}[t]$ (see Fig. 1), denoted by $C_{SR}(\theta_1)$, i.e.,

$$
A_S(\theta_1) = C_{SR}(\theta_1). \quad (8)
$$

Similarly, there exists a certain $\theta_2$ such that the effective bandwidth of $R_{SR}[t]$, denoted by $A_{SR}(\theta_2)$, is equal to the effective capacity of $R_{RD}[t]$, denoted by $C_{RD}(\theta_2)$:

$$
A_{SR}(\theta_2) = C_{RD}(\theta_2). \quad (9)
$$

Then, applying Eq. (7) we get

$$
\begin{aligned}
\Pr\{D_1 > x\} &\approx e^{-\theta_1 C_{SR}(\theta_1) x}; \\
\Pr\{D_2 > x\} &\approx e^{-\theta_2 C_{RD}(\theta_2) x}.
\end{aligned} \quad (10)
$$

Through Eq. (10), we further derive the probability density functions (pdf) of $D_1$ and $D_2$ by

$$
\begin{aligned}
g_1(x) &\approx \frac{d(1 - \Pr\{D_1 > x\})}{dx} = \theta_1 C_{SR}(\theta_1) e^{-\theta_1 C_{SR}(\theta_1) x}; \\
g_2(x) &\approx \frac{d(1 - \Pr\{D_2 > x\})}{dx} = \theta_2 C_{RD}(\theta_2) e^{-\theta_2 C_{RD}(\theta_2) x}.
\end{aligned} \quad (11)
$$

Based on Eq. (11), we can obtain the CCDF of $D_{total}$ as follows:

$$
\begin{aligned}
&\Pr\{D_{total} > x\} \\
&= \Pr\{D_1 + D_2 > x\} \\
&= 1 - \Pr\{D_1 + D_2 \le x\} \\
&= 1 - \int_0^x g_1(y) \int_0^{x-y} g_2(z) dz dy \\
&= \frac{\theta_1 C_{SR}(\theta_1) e^{-\theta_2 C_{RD}(\theta_2) x} - \theta_2 C_{SR}(\theta_2) e^{-\theta_1 C_{RD}(\theta_1) x}}{\theta_1 C_{SR}(\theta_1) - \theta_2 C_{SR}(\theta_2)}.
\end{aligned} \quad (12)
$$

Based on the International Telecommunication Union (ITU) recommendation [18], $D_{th}$ typically needs to be smaller than about 250 ms, much larger than the scale of the time frame in practical systems, which typically varies from a few milliseconds to tens of milliseconds. Therefore, we are more interested in the delay-bound violation probability under relatively large delay bound. Then, we let $x \to \infty$ and obtain

$$\lim_{x \to \infty} \Pr\{D_{total} > x\} \approx e^{-\min\{\theta_1 C_{SR}(\theta_1), \theta_2 C_{SR}(\theta_2)\}x}, \quad (13)$$

where

$$\frac{\max\{\theta_1 C_{SR}(\theta_1), \theta_2 C_{SR}(\theta_2)\}}{|\theta_1 C_{SR}(\theta_1) - \theta_2 C_{SR}(\theta_2)|}. \quad (14)$$

Eq. (13) implies the delay-bound violation probability eventually will be determined only by the minimum between $\theta_1 C_{SR}(\theta_1)$ and $\theta_2 C_{SR}(\theta_2)$, which are the exponentially decaying speed of the violation probability against the delay-bound. In other words, the link with pooper delay performance determines the overall end-to-end delay performance. Then, from the resource allocation point of view, if we want to improve the end-to-end delay performance, we should allocate more wireless resources to the link with poorer delay performances. Following this principle, we need to guarantee

$$\theta_1 C_{SR}(\theta_1) = \theta_2 C_{SR}(\theta_2). \quad (15)$$

Note that Eq. (15) implies that the SR link and the RD link should have the same statistical delay-QoS performances. However, this does suggest equal resource allocation across the SR link and the RD link, which is explained as follows. As illustrated in Fig. 1, the service process $R_{SR}[t]$ for the SR link also serves as the arrival process for the RD link. If channel gain $h_1$ and $h_2$ are identically distributed and we equally allocate power to the two link, we can see that at the RD link's queue, the arrival process and the service process' average rates are equal. In such a case, the queue will be build up and the delay-bound violation probability will approach 1. Therefore, the power allocation needs to be performed in an *asymmetric* manner, where the RD link should get more wireless resources.

Under Eq. (15), we re-derive the CCDF for $D_{total}$ and then get

$$\Pr\{D_{total} > x\} = (1 + x\theta_1 C_{SR}(\theta_1)) e^{-\theta_1 C_{SR}(\theta_1)x}. \quad (16)$$

Thus, given the delay-bound $D_{th}$ and the maximum tolerable violation probability $\xi$, we need to guarantee

$$\theta_1 C_{SR}(\theta_1) \geq -\frac{1}{D_{th}}\left(1 + W\left(-\frac{\xi}{e}\right)\right), \quad (17)$$

where $W(\cdot)$ is known as the Lambert-W function [17], which is the inverse function of $Z(W) = We^W$.

## IV. QoS-Driven Power Allocation Over Two-Hop Wireless Links

Having analyzed the statistical delay-QoS for the two-hop wireless links, we next focus on how to use the minimum transmit power $\kappa_1$ and $\kappa_2$ such that Eq. (1) holds under the given traffic load $\bar{A}$. Then, we derive the efficient solution of $\kappa_1$ and $\kappa_2$ through solving the following optimization problem:

$$\min_{(\kappa_1,\kappa_2)} \kappa_1 + \kappa_2 \quad (18)$$

s.t.: 1). $C_{SR}(\theta_1, \kappa_1) = \bar{A}$;

2). $\theta_1 C_{SR}(\theta_1, \kappa_1) = \theta_2 C_{RD}(\theta_2, \kappa_2)$;

3). $\theta_1 C_{SR}(\theta_1, \kappa_1) \geq -\frac{1}{D_{th}}\left(1 + W\left(-\frac{\xi}{e}\right)\right)$;

4). $A_{SR}(\theta_2, \kappa_1) = C_{RD}(\theta_2, \kappa_2)$.

In the above problem, all related effective capacities and effective bandwidths are functions of $\kappa_1$ and $\kappa_2$, respectively, and thus we write $\kappa_1$ and $\kappa_2$ as the corresponding independent variables in these functions. Moreover, constraint 1) results from Eq. (8) and plugging $R_S[t] = \bar{A}$ (see Section II) into Eq. (5); constraint 2) is required by Eq. (15); constraint 3) follows Eq. (17); constraint 4) is required by Eq. (9).

Before getting into details, we plug Eq. (2) into Eq. (5) and obtain

$$C_{SR}(\theta_1, \kappa_1) = -\frac{1}{\theta_1} \log\left(e^{\frac{1}{\kappa_1}} \kappa_1^{-\beta_1} G\left(1-\beta_1, \frac{1}{\kappa_1}\right)\right), \quad (19)$$

$$A_{SR}(\theta_2, \kappa_1) = \frac{1}{\theta_2} \log\left(e^{\frac{1}{\kappa_1}} \kappa_1^{\beta_1} G\left(1+\beta_2, \frac{1}{\kappa_1}\right)\right), \quad (20)$$

and

$$C_{RD}(\theta_2, \kappa_2) = -\frac{1}{\theta_2} \log\left(e^{\frac{1}{\kappa_2}} \kappa_2^{-\beta_2} G\left(1-\beta_2, \frac{1}{\kappa_2}\right)\right), \quad (21)$$

where $\beta_i \triangleq BT\theta_i$ for $i = 1, 2$, and

$$G(a,z) \triangleq \int_z^\infty t^{a-1}e^{-t}dt$$

is the upper incomplete Gamma function,

Note that constraint 3) lower-bounds the exponentially-decreasing speed of the delay-bound violation probability against the delay bound. The larger $\theta_1 C_{SR}(\theta_1, \kappa_1)$ is, the more power is required. Therefore, the minimum power to satisfy the delay QoS requirements (see Eq. (1)) is achieved when equality in constraint 3) hold. Following this principle, the optimal solution $(\kappa_1^*, \kappa_2^*)$ and the associated exponent $(\theta_1^*, \theta_2^*)$ can be derived through the following two steps:

**Step 1:** Solve for $\theta_1 = \theta_1^*$ and $\kappa_1 = \kappa_1^*$ such that

$$\begin{cases} C_{SR}(\theta_1, \kappa_1) = \bar{A}; \\ \theta_1 C_{SR}(\theta_1, \kappa_1) = -\frac{1}{D_{th}}\left(1 + W\left(-\frac{\xi}{e}\right)\right). \end{cases} \quad (22)$$

Eq. (22) implies that

$$\theta_1^* = -\frac{1}{\bar{A}D_{th}}\left(1 + W\left(-\frac{\xi}{e}\right)\right). \quad (23)$$

Then, we can obtain $\kappa_1^*$ through numerically solving

$$C_{SR}(\theta_1^*, \kappa_1) = \bar{A}.$$

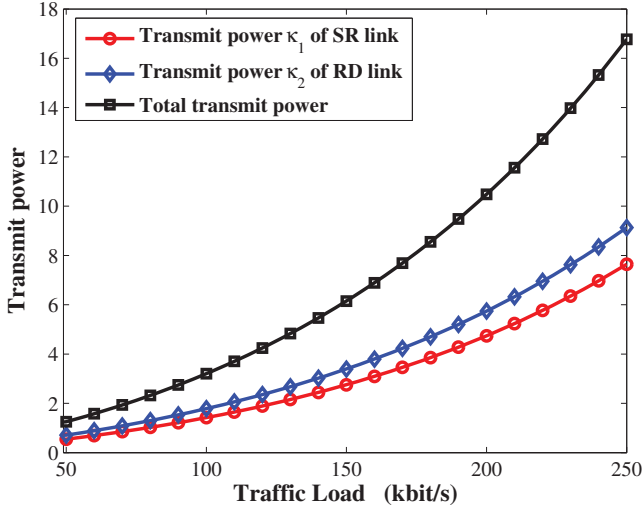
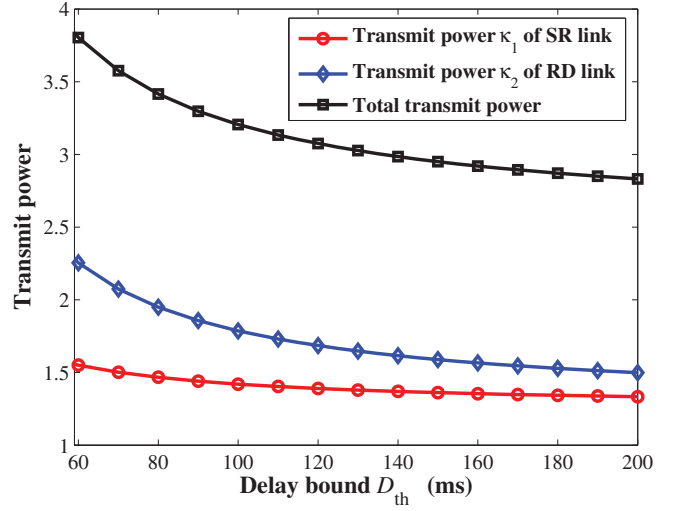

Fig. 3. Transmit power versus traffic load $\overline{A}$, where $D_{th} = 100$ ms, $\xi = 10^{-6}$, $E\{h_1\} = E\{h_2\}=1$, and $d_1 = d_2 = 50$ m.

Fig. 4. Transmit power versus delay bound $D_{th}$, where $\overline{A} = 100$ Kbps, $\xi = 10^{-6}$, $E\{h_1\} = E\{h_2\}=1$, and $d_1 = d_2 = 50$ m.

***Step 2:*** Solve for $\theta_2 = \theta_2^*$ and $\kappa_2 = \kappa_2^*$ such that

$$A_{SR}(\theta_2, \kappa_1^*) = C_{RD}(\theta_2, \kappa_2);$$
$$\theta_2 C_{RD}(\theta_2, \kappa_2) = -\frac{1}{D_{th}}\left[1 + W\left(-\frac{\xi}{e}\right)\right]. \quad (24)$$

Comparing the two equations in Eq. (24), We can first identify the $\theta_2 = \theta_2^*$ to satisfy

$$\theta_2 A_{SR}(\theta_2, \kappa_1^*) = -\frac{1}{D_{th}}\left[1 + W\left(-\frac{\xi}{e}\right)\right]. \quad (25)$$

It is easy to verify that $\theta_2 A_{SR}(\theta_2, \kappa_1^*)$ is an increasing function of $\theta_2$. Thus, we can conveniently obtain the unique $\theta_2^*$ through numerical searching method. Having $\theta_2^*$ and plugging it into the first equation of Eq. (24), we then search for the unique $\kappa_2 = \kappa_2^*$ to guarantee

$$A_{SR}(\theta_2^*, \kappa_1^*) = C_{RD}(\theta_2^*, \kappa_2). \quad (26)$$

## V. NUMERICAL RESULTS AND PERFORMANCE ANALYSES

We use numerical results to analyze the performances of our proposed QoS-driven power allocation scheme and investigate the impacts of delay-QoS requirements, traffic load, and the position of relay nodes on the power allocation. Throughout the simulations, we set $T = 2$ ms and $B = 10^5$ Hz. The source node, relay node, and destination node are located in a straight line, as depicted in Fig. 1. The distance between the source node and destination node is equal to 100 m. Without loss of generality, we set $E\{h_i\} = 1$ if the distance between the transmitting node and the receiving node is equal to 50 m, where $E\{\cdot\}$ denotes the expectation. Then, the expectation of $E\{h_i\}$ can be written as $(d_i/50)^{-\eta}$, where $\eta$ is the path loss exponent and typically varies from 2 to 6 for indoor environments without line-of-sight (LOS) [19]. For the numerical evaluations in this section, we set $\eta = 3$.

Figure 3 plots the consumed transmit power versus the traffic load under the specified delay QoS requirements with $d_1 = d_2$. We can see from Fig. 3 that the transmit power used for both SR link and RD link increase as the traffic load gets larger. However, the power of the RD link increases faster than that of the SR link. Fig. 4 also shows that the RD link needs more power as compared to the SR link. This implies that although the wireless channels for SR and RD links are symmetric, the power allocation over the two links is in an *asymmetric* manner.

Figures 4 and 5 plot the dynamics of the consumed transmit power for the two-hop relay links as functions of the delay QoS requirements. Figures 4 and 5 suggest that as the delay-QoS constraints gets more stringent, where either the delay bound $D_{th}$ gets smaller or the maximum tolerable delay-bound violation probability becomes smaller, more power needs to be used. Also, the RD link needs more power resources. In addition, as the delay-QoS constraints become looser, the difference between the power consumption of the SR link and that of the RD link gradually gets smaller.

Figure 6 illustrates the total consumed power as a function of the distance between the source node and relay node. Based on Figure 6, it is expected that the minimum power consumptions for QoS guarantees can be achieved when the relay node is not too close to either the source node nor the destination node. However, the optimal location for the relay node is not the center point ($d_1 = 50$ m) between the source and destination nodes. In contrast, the optimal location is closer to the destination node, which further demonstrates the optimality of the asymmetric power allocation scheme over the two-hop wireless relay links in terms of statistical delay-QoS provisioning.

## VI. CONCLUSIONS

We proposed the QoS-driven power-allocation scheme over two-hop wireless relay links to statistically upper-bound the

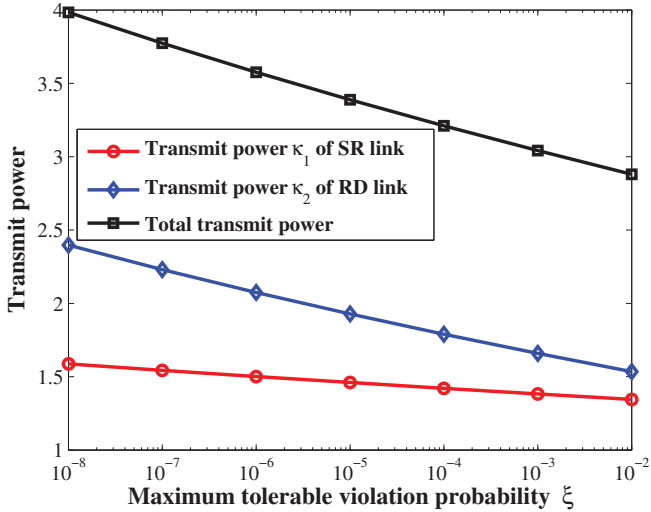

Fig. 5. Transmit power versus maximum delay-bound violation probability $\xi$, where $\overline{A}$ = 100 Kbps, $D_{th}$ = 70 ms, $E\{h_1\}$ = $E\{h_2\}$=1, and $d_1$ = $d_2$ = 50 m.

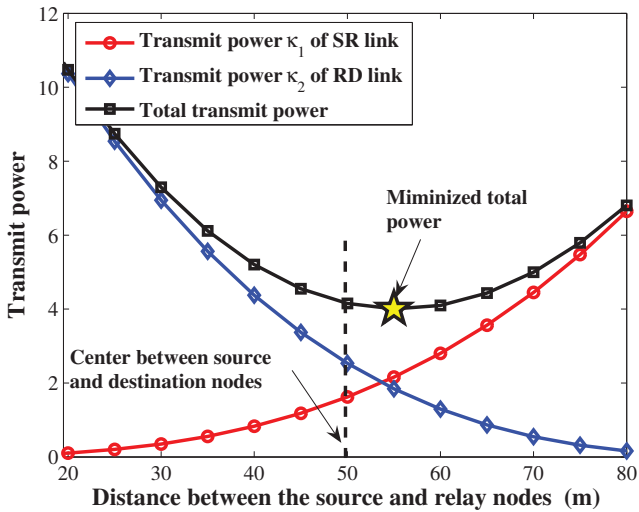

Fig. 6. Transmit power versus the distance $d_1$ between the source and relay nodes, where $\overline{A}$ = 100 Kbps, $D_{th}$ = 50 ms, and $\xi$ = $10^{-6}$.

*end-to-end* delay under the decode-and-forward relay transmissions. Specifically, by applying the effective capacity and effective bandwidth theories, we analyzed the delay-bound violation probability over two-top transmissions. We showed that the efficient approach for statistical-delay guarantees is to allocate the resources asymmetrically over the two-hop links, where the link closer to the destination node needs to get more resources. We also presented a set of simulations results to show the impact of the QoS requirements, traffic load, and position of the relay node on resource allocations.